\documentclass[aps,preprintnumbers,floatfix,amsmath,amssymb,11pt]{revtex4}

\usepackage{graphicx}
\usepackage{dcolumn}
\usepackage{bm}
\usepackage{verbatim}

\begin{document}

\title{Fractionally charged Wilson loops as a probe of $\theta$-dependence\\ in $CP^{N-1}$ sigma models:
Instantons vs. large N}
\author{ Patrick Keith-Hynes and H.B.~Thacker$^{}$ }
\affiliation{
 $^{}$Department of Physics, University of Virginia, Charlottesville, VA 22901 \\ }
\date{\today}


\begin{abstract}
The behavior of Wilson loops with fractional charge is used to study the $\theta$-dependence of the
free energy density $\varepsilon(\theta)$ for the $CP^1$, $CP^5$, and $CP^9$ sigma models in two spacetime dimensions. The function $\varepsilon(\theta)$
is extracted from the area law for a Wilson loop of charge $q=\theta/2\pi$. For $CP^1$, $\varepsilon(\theta)$
is smooth in the region $\theta\approx\pi$ and well-described by a dilute instanton gas throughout the range
$0<\theta<2\pi$. For $CP^5$ and $CP^9$ the energy exhibits a clear cusp and evidence for discrete, degenerate vacua
at $\theta = \pi$, as expected from large $N$ arguments. For $CP^9$ the $\theta$-dependence is in good 
quantitative agreement with the leading order large $N$ prediction $\varepsilon(\theta)=\frac{1}{2}\chi_t\theta^2$
throughout the range $0<\theta<\pi$.

\end{abstract}
\maketitle


\section{Introduction}

The structure of topological charge in the QCD vacuum is central to an understanding of low-energy
hadron dynamics. Lattice calculations have provided quantitative confirmation of the Witten-Veneziano \cite{Witten-Veneziano} relation
between the topological susceptibility of quenched QCD and the mass of the $\eta'$ meson \cite{Vicari}. 
The success of the Witten-Veneziano relation, which relates the topological susceptibility
to the chiral structure of the quenched $\eta'$ correlator \cite{Bardeen} is most easily understood 
in the context of the large $N$ approximation.
(In this case, the large $N$ approximation provides a justification for resumming quenched ``hairpin'' insertions.)
The observation of extended, coherent, codimension 1 sheets of topological charge in 
Monte Carlo generated SU(3) gauge configurations \cite{Horvath03,Ilgenfritz}
points to a new paradigm for the QCD vacuum. Contrary to the standard instanton liquid picture, the
vacuum is apparently a ``topological sandwich'' consisting of alternating sign membranes of topological charge.
Longstanding arguments of Witten \cite{Witten79} based on large-N chiral dynamics show that, for sufficiently
large N, instantons should disappear from the QCD vacuum and be replaced by codimension-1 membranes which are
in fact domain walls between discrete ``k-vacua,'' where the effective local value of $\theta$ differs from 
that of the true ground state by $2\pi k$, with $k$ an integer. The topological sandwich picture which has emerged from the 
Monte Carlo calculations is quite compatible with these large N arguments. This picture is in some ways
a 4-dimensional analog of Coleman's picture of $\theta$-dependence in the massive Schwinger model
\cite{Coleman}. In that model there are no instantons, and $\theta$ can be interpreted as a background electric field. The domain walls are
just charged particles, whose world lines have codimension 1. These world lines
separate regions in which the electric flux differs by one unit, i.e. $\Delta\theta=\pm2\pi$. 
A similar interpretation of $\theta$-dependence applies to the two-dimensional $CP^{N-1}$ sigma models,
which also have a U(1) gauge invariance. Luscher \cite{Luscher78} clarified the analogy between 2D $U(1)$ theories
and 4D Yang-Mills theory by pointing out the similar role played by the Chern-Simons tensors in the 2D and 4D theories.
The Wilson loop in the $CP^{N-1}$ model is analogous to a ``Wilson bag'' in QCD, which is an integral 
of the CS tensor of the Yang-Mills field over a three-dimensional surface. From this viewpoint the Wilson loop 
$\oint A\cdot dl$ in 2D should
be interpreted as a one-dimensional surface integral of the Chern-Simons flux $\epsilon_{\mu\nu}A^{\nu}$. Monte Carlo studies of the $CP^{N-1}$ 
models \cite{Ahmad,Lian} have shown 
that for $N>3$ these models exhibit a topological structure quite analogous to that observed in lattice QCD,
with the vacuum occupied by extended coherent topological charge structures of codimension 1 \cite{Ahmad}. 
In striking contrast, the 
topological charge distributions in the $CP^1$ and
$CP^2$ models are found to be dominated by small instantons \cite{Lian}. The $CP^{N-1}$ 
models thus provide a detailed example
of Witten's arguments that instantons should ``melt'' or become irrelevant at large $N$. For 
the $CP^{N-1}$ models, the instanton melting point is found to be about $N\approx 4$ \cite{Lian}. 
(As discussed in \cite{Luscher_chit}, 
the value of $N$ at which instantons become irrelevant 
may depend on the lattice formulation, but for any latticization, at least the
$CP^1$ model is expected to be dominated by small instantons.)

Although the value of the $\theta$ parameter in real-world QCD is zero to high accuracy, a deeper understanding of
topological charge structure can by obtained by considering gauge theories with nonzero $\theta$ terms. 
Unfortunately, Monte Carlo studies of QCD at nonzero $\theta$ are severely hampered by the fact that the
$\theta$ term contributes an imaginary part to the Eucidean action. As a result, the exponentiated action in the
path integral cannot be interpreted as a probablility, precluding the direct simulation of such theories 
by Monte Carlo techniques. Naive reweighting methods, in which the $e^{i\theta\nu}$ factor in the path integral
is introduced in the ensemble average over $\theta=0$ configurations, are sufficient for probing small values
of $\theta$. But some of the most interesting issues associated with multi-phase structure are most directly
addressed by studying the region $\theta\approx\pi$, where reweighting methods are ineffective. 
More sophisticated techniques for extending the reach
of Monte Carlo techniques to large $\theta$ have been introduced \cite{Burkhalter,Azcoiti,Beard,Imachi}, 
but definitive results near $\theta\approx\pi$ are still difficult to obtain by these methods. 

In this paper we present a method for exploring $\theta$-dependence in two-dimensional $U(1)$ gauge theories
which is based on the calculation of fractionally charged Wilson loops in the $\theta=0$ theory. The results for 
the free energy density $\varepsilon(\theta)=E(\theta)/V$ in the $CP^{N-1}$
models exhibit the power of this method, providing direct numerical evidence for a
first order phase transition at $\theta=\pi$ for the $CP^5$ and $CP^9$ models. In clear contrast with the larger $N$ models, the instanton dominated
$CP^1$ model exhibits smooth behavior in the $\theta=\pi$ region, as expected from a dilute instanton
gas calculation. (Our results do not rule out a second order phase transition at $\theta=\pi$
in the $CP^1$ model, which is expected from theoretical arguments \cite{Haldane}).   

\section{Fractionally Charged Wilson Loops}

The central observation which we exploit here is that, in the path integral for a 2D U(1) gauge theory, 
including a closed Wilson
loop with charge $q=\theta/2\pi$ is equivalent to including a $\theta$ term in the two-volume
enclosed by the loop. The Wilson loop goes around the boundary $\partial V$ of a two-volume $V$,
\begin{equation}
\label{eq:thetaterm}
\theta \int_V d^2x Q(x) = \frac{\theta}{2\pi}\oint_{\partial V} A\cdot dx
\end{equation}
where 
\begin{equation}
Q(x) = \frac{1}{2\pi}\epsilon_{\mu\nu}F^{\mu\nu}
\end{equation}
is the topological charge density.
Since the $CP^{N-1}$ models have nonzero topological susceptibility, the free energy per unit volume $\varepsilon(\theta)$ inside
the loop is greater than $\varepsilon(0)$ outside. This gives rise to a linear confining potential between fractional charges 
or equivalently, an area law for large Wilson loops. The coefficient of the area law determines the difference
in vacuum energy inside and outside the loop:
\begin{equation}
\langle W_{\cal C}(q)\rangle \sim \exp\left[-\left(\varepsilon(\theta)-\varepsilon(0)\right)V\right]|_{\theta=2\pi q}
\end{equation}
where ${\cal C}=\partial V$. 
Note that since $\varepsilon(\theta+2\pi)=\varepsilon(\theta)$, the coefficient of the area 
law is periodic in the charge $q$.
At integer values of charge, the confining force is completely screened and the area term in the Wilson
loop vanishes. The behavior observed in the region $q\approx 1/2$ distinguishes between an instanton-dominated
model and a large $N$ domain wall model. For an instanton theory, a dilute gas approximation gives
\begin{equation}
\label{eq:instanton_gas}
\varepsilon(\theta) -\varepsilon(0) = \chi_t(1-\cos\theta)
\end{equation}
where $\chi_t$ is the topological susceptibility. On the other hand, large N
considerations \cite{Witten79} predict quadratic behavior, with periodicity leading to cusps at odd
multiples of $\pi$,
\begin{equation}
\label{eq:largeN}
\varepsilon(\theta) - \varepsilon(0) = \frac{1}{2}\chi_t\; {\rm min}_{k\epsilon {\cal Z}}\;(\theta - 2\pi k)^2
\end{equation}
Physically, these cusps represent ``string-breaking'' or vacuum screening, which occurs when it is energetically favorable 
to screen the background electric flux by one unit from $\frac{1}{2}+\delta$ to $-\frac{1}{2}+\delta$. For large loops,
this can be interpreted as a first order phase transition taking place inside the loop.

\begin{figure}
\label{fig:largeN}
\begin{center}
\includegraphics{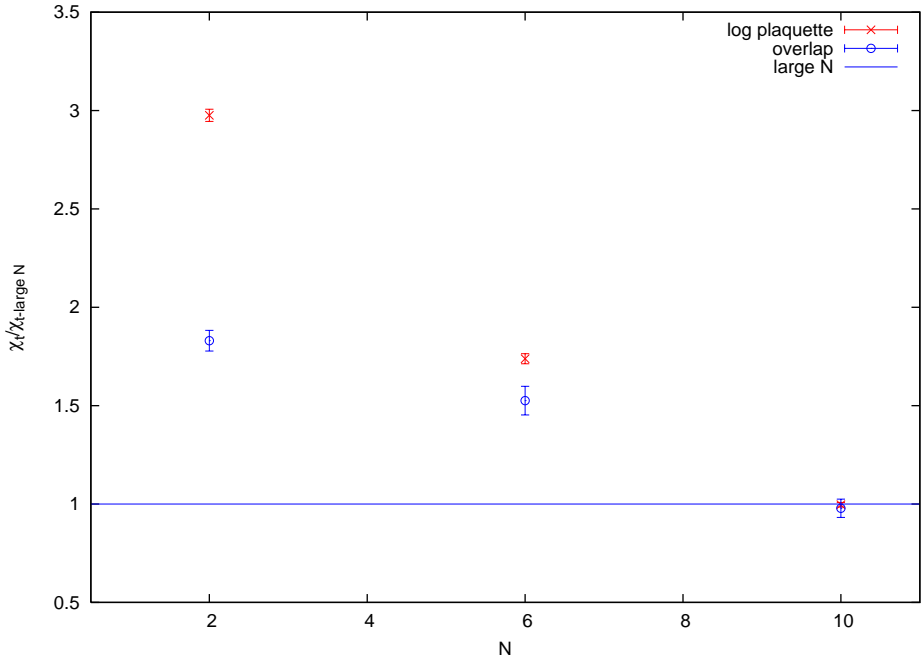}
\caption{Behavior of the topological susceptibility as a function of $N$, divided by the large N prediction,
for both log plaquette and overlap definitions of lattice topological charge.}
\end{center}
\end{figure}

We study the Wilson loops for the $CP^1$, $CP^5$, and $CP^9$ models. As shown in Ref. \cite{Lian},
$CP^1$ is instanton-dominated, while $CP^5$ and $CP^9$ are both above the instanton ``melting point,'' 
It was found that the topological
charge distribution for $CP^1$ is dominated, at large $\beta$, by small instantons, while for $CP^5$ and $CP^9$
instantons do
not occur but topological charge instead appears in the form of coherent one-dimensional domain wall-like structures.
In Ref. \cite{Lian} results for the topological suscepibility were compared with the leading-order large N 
prediction,
\begin{equation}
\chi_t = \frac{3\mu^2}{4\pi N}
\end{equation}
where $\mu$ is the mass gap (nonsinglet meson mass). As shown in Fig. 1, the result for $CP^5$ is still
significantly above the large $N$ prediction, but $CP^9$ is in good agreement. 
Fig. 1 plots the ratio of the measured value of $\chi_t$ to the 
large $N$ prediction (\ref{eq:largeN}). The $\times$'s and $o$'s denote the results for log plaquette and overlap
topological charge, respectively. For each of the $CP^{N-1}$ models, the value of the coupling constant $\beta$ was adjusted
to give a correlation length of $\approx 5$ to $7$. [The $CP^1, CP^5$, and $CP^9$ data were obtained at $\beta = 1.2,
0.9$, and $0.8$ respectively, for which the meson masses are $\mu = 0.179(3), 0.186(3),$ and $0.212(2)$.
Note that for $CP^5$ and $CP^9$, $\chi_t$ scales properly as $\chi_t\tilde \mu^2$, while for $CP^1$ the small instantons
cause $\chi_t/\mu^2$ to diverge in the continuum limit.]
The calculations were done on both $50\times 50$ and $100\times 100$ lattices and the effect of finite volume was found
to be negligible. 
The results plotted in Figure 1 indicate that by studying these three models, we cover
the entire range of topological charge dynamics from $CP^1$ which is instanton-dominated, to $CP^9$ where bulk topological
properties are described with reasonable accuracy by the large N approximation. This assertion is further supported by the
results for $\theta$ dependence of the vacuum energy presented in this paper.

\begin{figure}
\label{fig:2pi_plot}
\begin{center}
\includegraphics{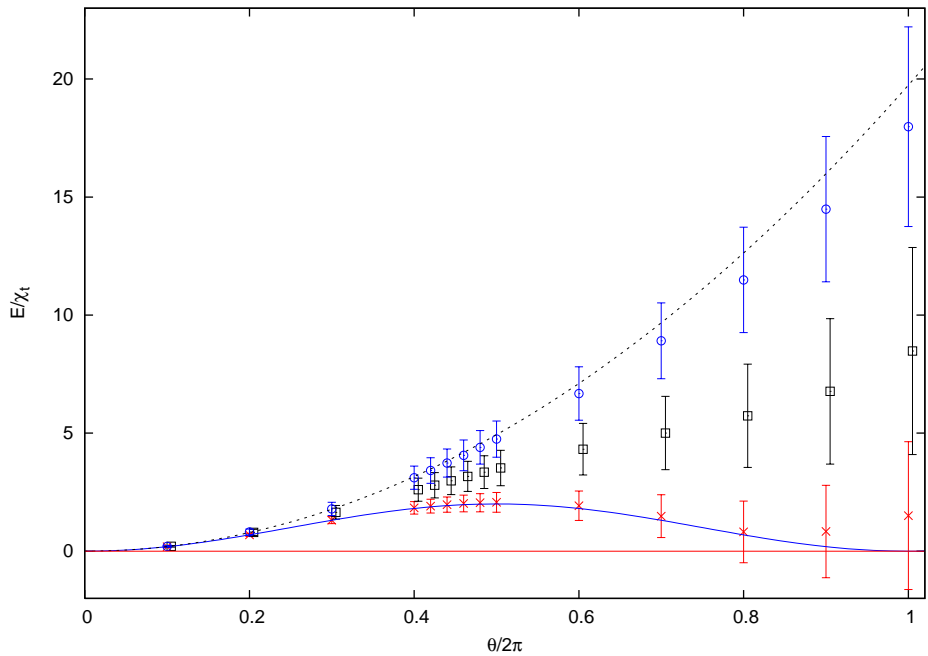}
\caption{The free energy density $\varepsilon(\theta)$ for $CP^1$ ($\times$'s), 
$CP^5$ ($\square$'s), and $CP^9$ ($O$'s) as measured from fractionally charged
Wilson loops. The lower and upper curves are the instanton gas 
and large $N$ predictions, normalized to the same topological susceptibility. Note that, for $\theta/2\pi>\frac{1}{2}$ the large N curve 
represents the energy of the false (unscreened) vacuum.}
\end{center}
\end{figure}

One of the most striking features of the Monte Carlo results for
fractionally charged Wilson loops is the difference between the behavior in $CP^1$ versus that in $CP^5$ 
and $CP^9$ in the region $\pi<\theta<2\pi$. Since $\varepsilon(\theta)$ is
periodic and an even function of $\theta$, its value in this region should be determined by its value in
the range $0<\theta<\pi$ by reflection around $\pi$,
\begin{equation}
\label{eq:reflection}
\varepsilon(\theta) = \varepsilon(2\pi-\theta)
\end{equation}
As seen in Figure 2, the measured value of $\varepsilon(\theta)$ for $CP^1$ ($\times$'s) is, within errors, nicely periodic and symmetric around $\theta=\pi$,
and in fact fits well to the dilute instanton gas prediction (\ref{eq:instanton_gas}) throughout the range $0<\theta<2\pi$.
On the other hand, for $CP^5$ ($\square$'s) and $CP^9$ (o's), as shown in Fig. 2, the coefficient extracted from a simple area-law fit to the Wilson 
loops continues to rise beyond $\theta=\pi$, violating the expected symmetry (\ref{eq:reflection}). 
We will argue that the behavior of $CP^5$ and $CP^9$ for $\theta>\pi$ is
an effect of having two nearly degenerate ground states.
This behavior can be easily understood in terms of the large N picture\cite{Witten79}, in which there are
two nearly degenerate quasi-vacua when $\theta\approx\pi$. One vacuum has a background electric 
field $\theta/2\pi\approx +\frac{1}{2}$. The other is the one in which a unit of flux has been screened,
so that $\theta/2\pi\approx -\frac{1}{2}$.
A Wilson loop with length $R$ in the spatial direction and $T$ in the time direction
can be interpreted as the $T$-dependent propagator of a ``string'' of length $R$, consisting of a $+q$ and a $-q$ 
charge with an amount $q=\theta/2\pi$ of electric flux between them. This state has a large overlap with the 
vacuum state containing background flux of $\theta/2\pi$. But for $\theta>\pi$, the true ground state is the one
where the flux has been screened by one unit to $\theta/2\pi-1$. In order for the Wilson line to couple
to this screened vacuum, the flux string must break via vacuum polarization. It is thus expected that, 
for $\theta>\pi$, the Wilson line will have a much larger overlap with the false (unscreened) vacuum than
with the true (screened) vacuum. As a result, the Wilson loop area law tends to be determined by the energy of the
unscreened vacuum, even for $\theta>\pi$ where the screened vacuum has lower energy.   
Since the Wilson line couples preferentially to the unscreened vacuum, we expect that our results for $\varepsilon(\theta)$
are measuring the true ground state energy throughout the range $0<\theta <\pi$, where the unscreened vacuum is the true vacuum.
By invoking the reflection symmetry (\ref{eq:reflection}) we obtain a complete determination of $\varepsilon(\theta)$.

\begin{figure}
\label{fig:plotCP1}
\begin{center}
\includegraphics{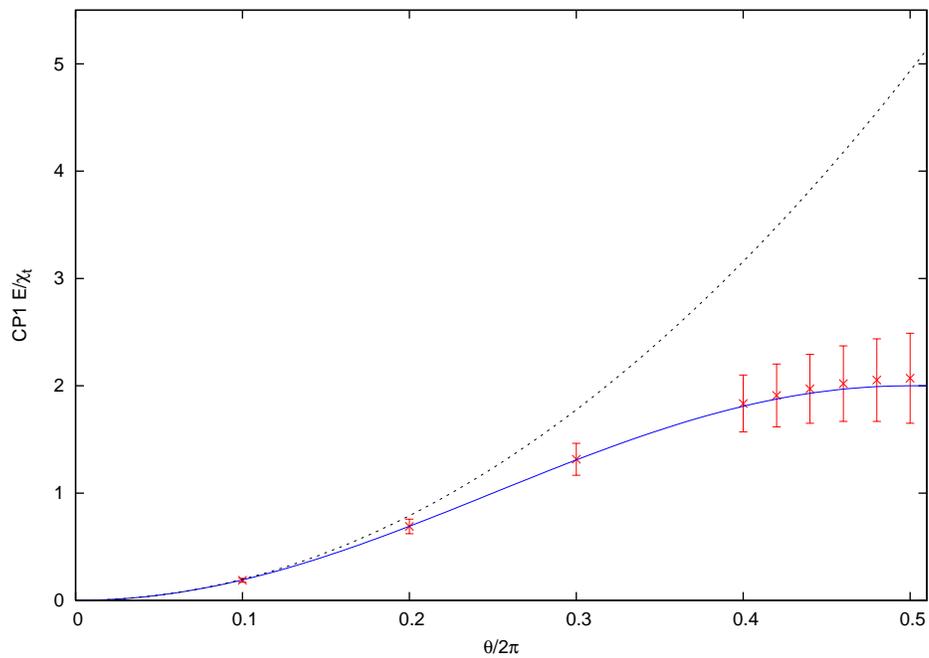}
\caption{$\varepsilon(\theta)$ for the $CP^1$ model.}
\end{center}
\end{figure}

\begin{figure}
\label{fig:plotCP5}
\begin{center}
\includegraphics{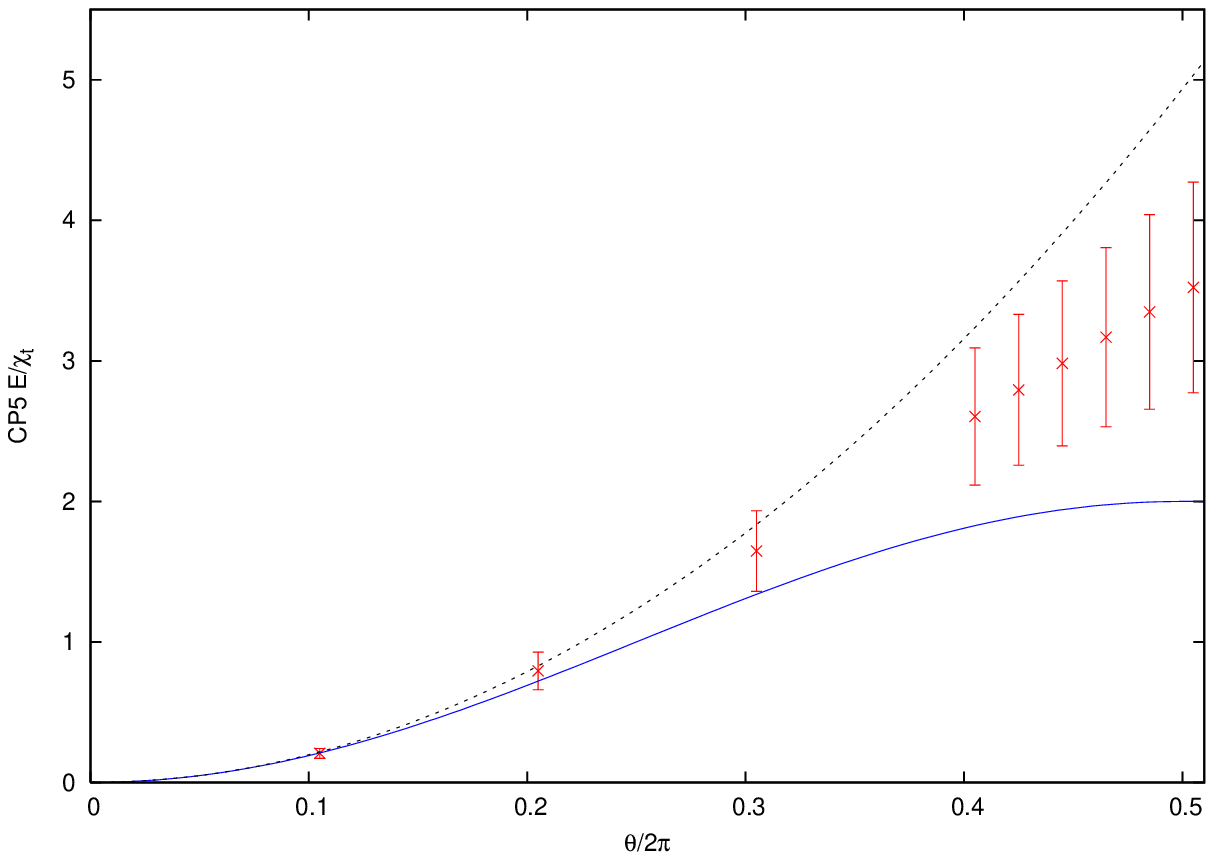}
\caption{$\varepsilon(\theta)$ for the $CP^5$ model.}
\end{center}
\end{figure}

\begin{figure}
\label{fig:plotCP9}
\begin{center}
\includegraphics{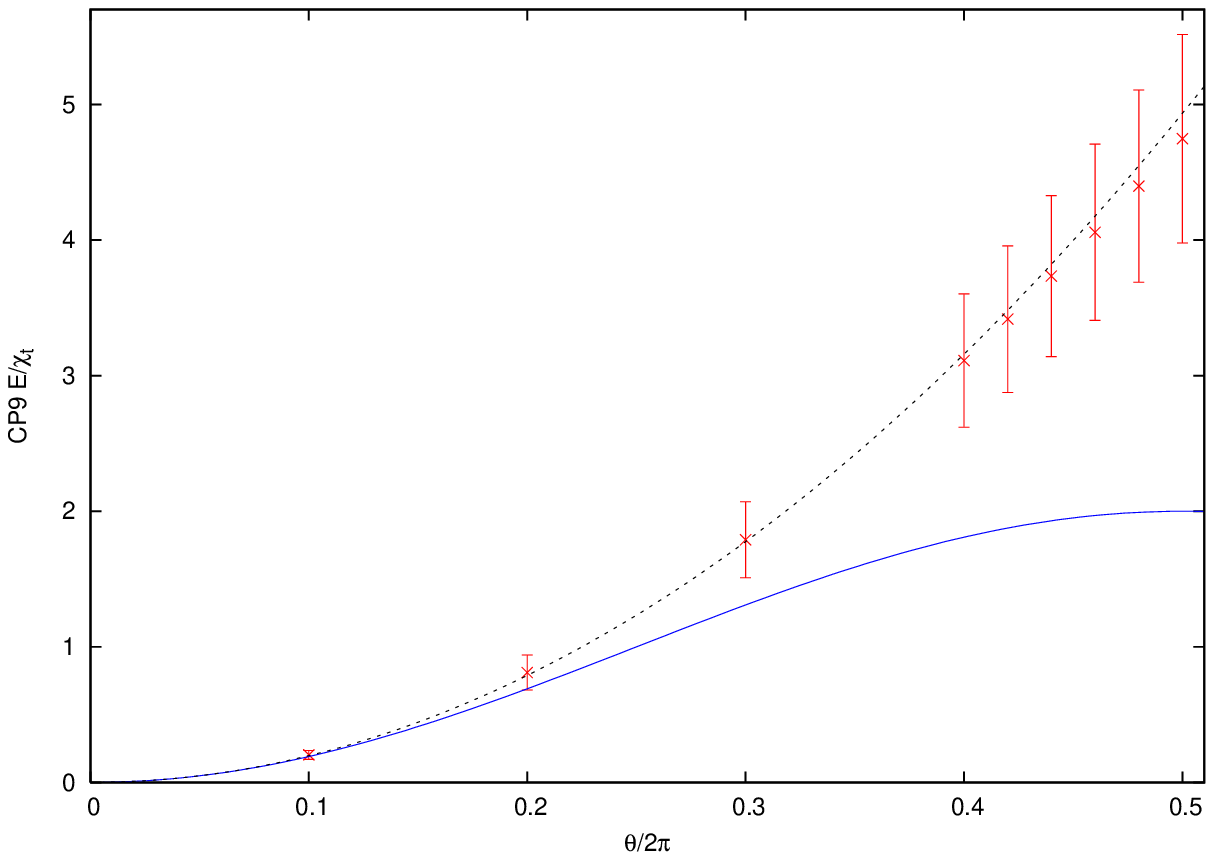}
\caption{$\varepsilon(\theta)$ for the $CP^9$ model.}
\end{center}
\end{figure}

The results for $\varepsilon(\theta)$ for $CP^1$ are plotted for $0<\theta<\pi$ in Fig. 3. With the topological susceptibility fixed to the
value obtained from the fluctuation of the integer-valued global topological charge, $\chi_t=\langle\nu^2\rangle/V$, 
the solid line is a zero-parameter fit to the dilute instanton gas formula (\ref{eq:instanton_gas}).
Also plotted (dotted line) is the leading order large N prediction,
\begin{equation}
\varepsilon(\theta) - \varepsilon(0) = \frac{1}{2}\chi_t \theta^2
\end{equation}
The corresponding results for $CP^5$ and $CP^9$ are shown in Figs. 4 and 5. The solid and dotted curves are again the instanton gas 
and large $N$ predictions with
normalization fixed to the measured $\chi_t$. 
Unlike the $CP^1$ case, the data for $CP^5$ and $CP^9$ shows a clear departure from the instanton gas model in the direction of
large N.
In particular, there is convincing evidence that $\varepsilon(\theta)$ has a positive slope at $\theta=\pi$, and since 
$\varepsilon(\theta)=\varepsilon(2\pi-\theta)$, this implies a cusp with discontinuous derivative at $\theta=\pi$. 
Moreover, the results for $\epsilon(\theta)$ for the $CP^9$ model (Fig. 5) are in good agreement with the large N 
prediction (\ref{eq:largeN}) in both the magnitude of $\chi_t$ and in the shape of the function $\epsilon(\theta)$.
(The $\chi_t$ used here is from the log plaquette charge, but for $CP^9$ there is little difference between the log plaquette
and overlap values for $\chi_t$. See Fig. 1.)
In all of the models, the value of $\chi_t$ obtained from fractionally charged Wilson loops using 
$\chi_t = \varepsilon''(0)$ is in excellent agreement with that obtained from the fluctuations of
the global topological charge $\chi_t=\langle\nu^2\rangle/V$. This agreement provides additional evidence 
that the fractionally charged Wilson loop method gives an accurate determination of $\varepsilon(\theta)$
over the entire range $0<\theta<\pi$. 

\section{Lattice $CP^{N-1}$ Models}

The Lagrangian for $CP^{N-1}$ sigma models in the continuum is
\begin{equation}
\label{eqn:L_cpn1}
L = \partial_{\mu}z_i^*\partial_{\mu}z_i + (z_i^*\partial_{\mu}z_i)(z_j^*\partial_{\mu}z_j)
\end{equation}
where $z^i$ are N complex fields, $i=1,\dots,N$, satisfying a constraint $z_{i}^{*}z_{i}=1$.
This Lagrangian is invariant under a local U(1) gauge transformation: $z_{i}(x)\to e^{ia(x)}z_{i}(x)$, 
for arbitrary space-time dependent $a(x)$. We can introduce a dummy gauge field $A_{\mu}$ and 
rewrite the Lagrangian as
\begin{equation}
\label{eqn:L_cpn2}
L = (\partial_{\mu} - iA_{\mu})z_{i}^{*}(\partial_{\mu} + iA_{\mu})z_i
\end{equation}
where 
\begin{equation}
\label{A_mu_cpn0}
A_{\mu} = \frac{1}{2}i \left[z_i^{\dagger} (\partial_{\mu}z) - (\partial_{\mu}z_i^{\dagger})z_i\right]
\end{equation}

To put $CP^{N-1}$ models on the lattice, we introduce $U(1)$ link fields $U(x, x+\hat{\mu}) = e^{iA_{\mu}(x)}$.
$CP^{N-1}$ fields are defined on the sites as $z_i(x)$. The lattice action 
consists of gauge-invariant nearest-neighbor hopping terms,
\begin{equation}
\label{eqn:cpn_action}
S = \beta N \sum_{x, \hat{\mu}}z_i(x)^\dagger U(x, x+\hat{\mu})z_i(x+\hat{\mu}) + c.c.
\end{equation}
This lattice action is used in our Monte Carlo simulation to generate an ensemble of field configurations. 
The $z_i$ fields are updated by a Cabibbo-Marinari heat bath algorithm (i.e. by applying an $SU(2)$ heat bath to all pairs of 
components $z_i$, $z_j$), while $U(1)$ link fields are updated by 
a multi-hit Metropolis algorithm.

In the continuum theory, the topological charge density is proportional to the electric field strength,
\begin{equation}
Q(x) = \frac{1}{2\pi}\epsilon_{\mu\nu}\partial^{\mu}A^{\nu}
\end{equation}
The simplest definition of the local topological charge density on the lattice is thus obtained from the plaquette phase,
\begin{equation}
\label{plaqcharge}
Q_{\nu} = \frac{-i}{2\pi} \ln P_{\nu}
\end{equation}
In recent studies of topological charge structure in the $CP^{N-1}$ models\cite{Ahmad,Lian} (as well as in QCD \cite{Horvath03}), it was found that 
a ``fermionic'' construction of the lattice topological charge density based on the overlap Dirac operator
provides a much clearer view of coherent structures (both domain walls and instantons). The density obtained
from the plaquette phase has a high level of short range anticorrelated noise which is smoothed out in the
overlap-based topological charge distribution. Nevertheless, the two choices of lattice topological charge give similar results
for bulk topological susceptibility, particularly for larger $N$ (see Fig. 1).

The Wilson loops we consider in this paper can be interpreted as world lines of fractionally charged test particles. 
In the compact formulation we are using, the link phase associated with a particle of charge $q$ is $U^q$, a fractional
power of the link variable. In general this would require a branch choice which will not be gauge invariant.
However, for $U(1)$ gauge theories in two space-time dimensions, there is a straightforward gauge invariant procedure for
calculating closed Wilson loops with fractional charge, using the lattice Stoke's theorem,
\begin{equation}
\label{eq:fracloop}
\langle {\cal W}(C)\rangle = \langle\prod_{l\epsilon C}U_i^q\rangle = \exp{q\sum_{\nu\epsilon V} \ln P_{\nu}}
\end{equation}
where the loop $C$ is the boundary of the enclosed volume $V$, $l$ labels the links on the loop and $\nu$
labels the plaquettes enclosed. Here $P_{\nu}$ is the gauge invariant product of links around a single plaquette, 
with its log taken to be on the principal branch,
\begin{equation}
-\pi < -i\ln P_{\nu} \leq \pi
\end{equation}
(Note that this branch choice is gauge independent, unlike a branch choice for the link phase $U^q$.) 

\begin{figure}
\label{fig:static0.3_cp1}
\begin{center}
\includegraphics{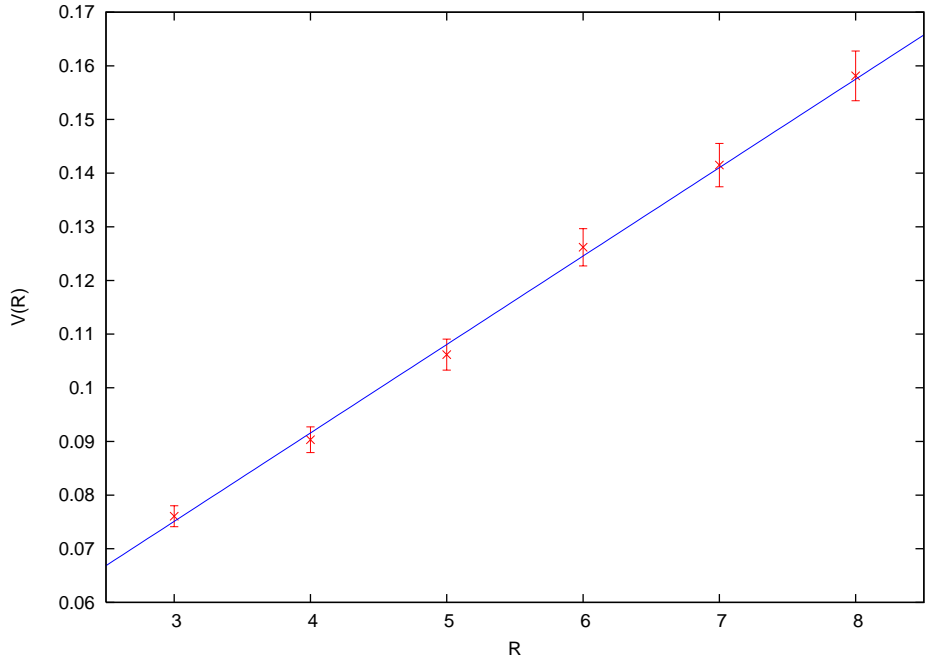}
\caption{Static potential for the $CP^1$ model with Wilson loop charge $q=0.3$.}
\end{center}
\end{figure}

\begin{figure}
\label{fig:static0.3_cp5}
\begin{center}
\includegraphics{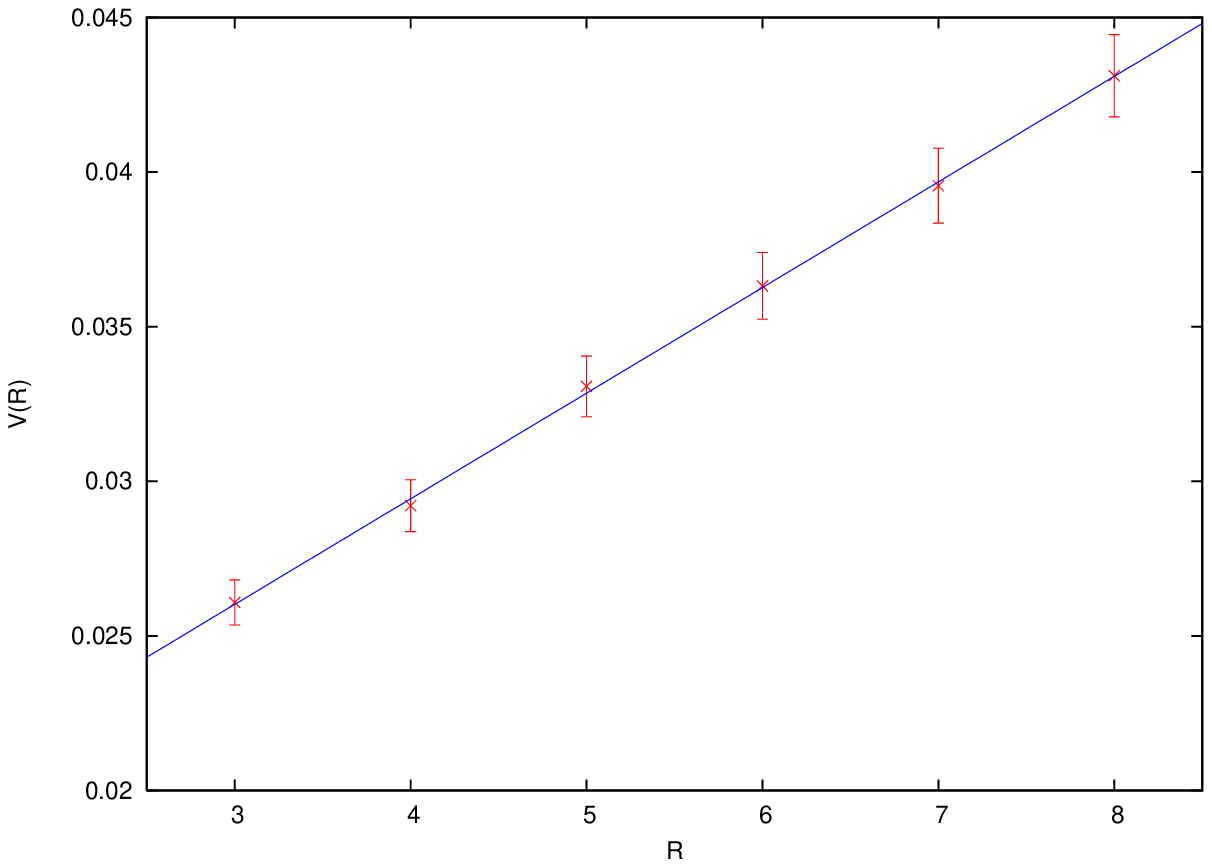}
\caption{Static potential for the $CP^5$ model with Wilson loop charge $q=0.3$.}
\end{center}
\end{figure}

\begin{figure}
\label{fig:static0.3_cp9}
\begin{center}
\includegraphics{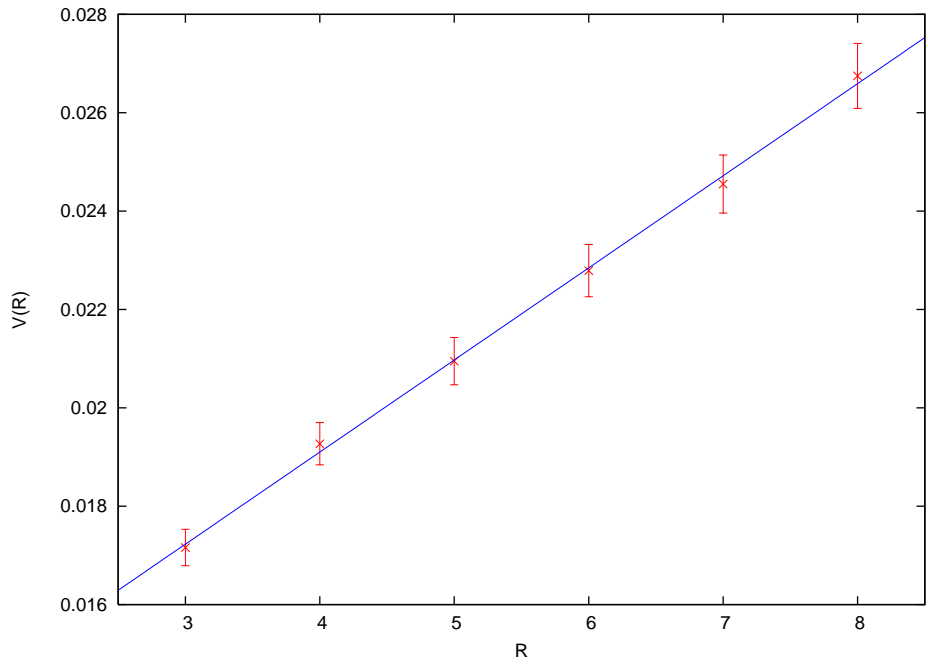}
\caption{Static potential for the $CP^9$ model with Wilson loop charge $q=0.3$.}
\end{center}
\end{figure}

To extract the ground state energy density $\varepsilon(\theta)$ from the Wilson loop expectation values,
we consider rectangular loops ${\cal W}_q(R,T)$ with sides $R$ and $T$ and charge $q$.
The fitting of the Wilson loops for charges $q<\frac{1}{2}$ is particularly simple in that 
there is very little deviation from a pure area law, even for loops as small as $2\times 2$.
This allows a surprisingly accurate determination of $\varepsilon(\theta)$ from the area law coefficient
over the entire range $0<\theta<\pi$.
It is convenient to extract the area law coefficient by first studying the static potential $V_q(R)$. 
Regarding T as Euclidean time, the static potential is calculated by fitting the data 
at each fixed value of $R$ to an exponential $T$ dependence,
\begin{equation}
\label{eq:expfit}
\langle {\cal W}_q(R,T)\rangle \sim {\rm const.}\times \exp(-V_q(R)T)
\end{equation}
The $T$-dependence for any fixed $R$ is well described by a single exponential for 
$T>1$. Moreover, the static potential $V_q(R)$ is found to be quite accurately linear,
\begin{equation}
V_q(R)\sim R\times \varepsilon(2\pi q)
\end{equation}
The static potential for $q = 0.3$ is shown in Figs. 6, 7, and 8 for $CP^1$, $CP^5$, and $CP^9$, respectively.
The value of $\varepsilon(\theta)$ is determined by fitting the slope of the linear potential.

As we have discussed, for the $CP^1$ model, the function $\epsilon(\theta)$ decreases for $\theta>\pi$ and
is nicely consistent with the expected symmetry (\ref{eq:reflection}). At $\theta=2\pi$, the slope of the linear potential
for $CP^1$ is consistent with zero, as seen in Fig. 9.  For the $CP^5$ and $CP^9$ models,
the behavior for $\theta>\pi$ is quite different. As we have discussed, the measured value of
$\epsilon(\theta)$ continues to increase beyond $\theta=\pi$ and violates reflection symmetry around $\pi$.
The explanation of this behavior in terms of coupling to the unscreened false vacuum suggest that, for $q>\frac{1}{2}$, we might be able 
to see evidence of the true vacuum by observing a levelling off of the slope of the linear potential 
for very large values of $R$ (and $T$) as the effect of vacuum screening sets in. Within the limits of our
statistics, we have not been able to see significant evidence for a change of slope in $V(R)$, 
even for loops as large as $10\times 10$. 
Figs. 9 and 10 show the results for $V(R)$ with a Wilson loop charge of 1.0. 
We see that the $CP^5$ and $CP^9$ results in Fig. 10 still
indicate a linear confining potential. The situation here is analogous to the
problem of observing string breaking in QCD, i.e. the screening of the linear potential by closed quark
loops. The direct observation of string breaking from Wilson loops
in full QCD is notoriously difficult (c.f. \cite{Schilling}), and most studies of this process have had to resort to coupled channel 
calculations. 

\begin{figure}
\label{fig:static1.0_cp1}
\begin{center}
\includegraphics{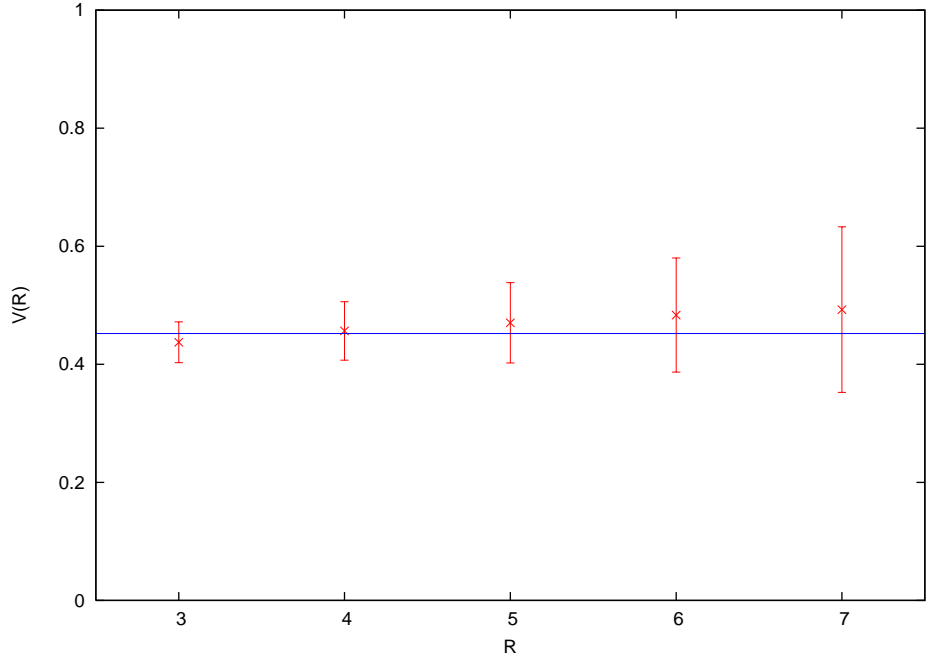}
\caption{Static potential for the $CP^1$ model with Wilson loop charge $q= 1.0$. The solid line is the best zero-slope fit. }
\end{center}
\end{figure}

\begin{figure}
\label{fig:static1.0_cp59}
\begin{center}
\includegraphics{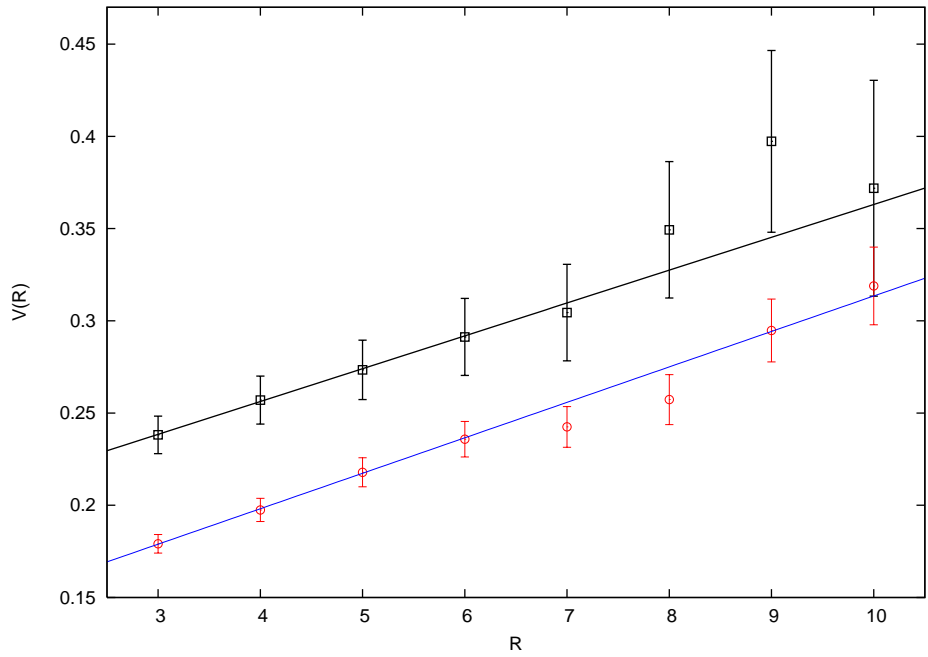}
\caption{Static potential for the $CP^5$ and $CP^9$ models with Wilson loop charge $q=1.0$.}
\end{center}
\end{figure}

\section{Instantons vs. large N}

The $\theta$ dependence of the free energy that we have calculated from fractionally charged Wilson loops
provides a detailed view of the transition between instanton physics and large-N physics
in an asymptotically free gauge theory. The ability to obtain accurate results for the free energy as 
a function of $\theta$ over the whole range $0<\theta<\pi$ opens new possibilities for the Monte Carlo 
exploration of topological phase structure in two-dimensional gauge theories. The understanding of this phase structure
has broad implications, not only as an analog of QCD vacuum structure in four dimensions, but also for
the structure of topologically ordered states in two-dimensional superconductors \cite{Sondhi}. The appearance of
a cusp in $\epsilon(\theta)$ at $\theta=\pi$ indicates the presence of a first order phase transition, 
pointing to the existence of two discrete vacuum states which are crossing in energy at $\theta=\pi$, as first 
suggested by Witten \cite{Witten79} for both QCD and $CP^{N-1}$ models. The long range coherent 
codimension 1 topological charge membranes which have been observed
in Monte Carlo configurations \cite{Horvath03,Ahmad} are naturally interpreted as domain walls between discrete vacua.

It is interesting to consider the calculation of $\varepsilon(\theta)$ in the dilute instanton gas approximation 
(\ref{eq:instanton_gas}) and compare it with the assumptions that lead to the large N prediction (\ref{eq:largeN}).
We write the Euclidean path integral with periodic boundary conditions as a sum over "winding number" 
(global topological charge) sectors,
\begin{equation}
{\cal Z} = \sum_{\nu=-\infty}^{\infty} Z_{\nu} e^{i\nu\theta}
\end{equation}
where $Z_{\nu}$ is the path integral over configurations with total topological charge $\nu$.
For a dilute instanton gas, this is given as a sum over instantons and antiinstantons,
\begin{equation}
\label{eq:sum_over_instantons}
Z_{\nu} = \sum_{\nu_1,\nu_2 =0}^{\infty}\frac{z^{\nu_1}}{\nu_1!}\frac{z^{\nu_2}}{\nu_2!}\delta_{\nu,\nu_1-\nu_2}
= I_{\nu}(2z)
\end{equation}
Here $z$ is the partition function for a single instanton or antiinstanton in a box of volume $V$.
Thus the partition function is
\begin{equation}
{\cal Z} = \sum_{\nu=-\infty}^{\infty}I_{\nu}(2z)e^{i\nu\theta} = \exp(2z\cos\theta)
\end{equation}
This gives the result (\ref{eq:instanton_gas}) for the free energy $\varepsilon(\theta)=-\ln{\cal Z}/V$. 

To contrast this result with the large N prediction (\ref{eq:largeN}) we go back to the sum
over topological charge sectors and employ a Poisson resummation,
\begin{equation}
\label{eq:Poisson}
{\cal Z} = \sum_{\nu}Z_{\nu}e^{i\nu\theta}
=\int dQ Z_Q e^{iQ\theta}\sum_{\nu}\delta(Q-\nu)
= \sum_{k=-\infty}^{\infty}\int dQ Z_Q e^{iQ(\theta - 2\pi k)}
\end{equation}
Here we assume that for large volume, the partition function over topological charge sectors $Z_{\nu}$
can be smoothly interpolated between integer values $Q=\nu$ by a continuous function $Z_Q$.
We can now show that the $\theta$ dependence (\ref{eq:largeN}) predicted by large N arguments
actually follows more generally from the assumption that the fluctuations of global topological charge 
are gaussian. In particular, the integer $k$ that labels discrete k-vacua is dual to the topological charge $\nu$
in the sense of Poisson resummation. If we take
\begin{equation}
Z_Q = Z_0e^{-Q^2/2\chi_t V}
\end{equation}
then the partition function can be written 
\begin{equation}
\label{eq:kvacua}
{\cal Z} = {\rm const.}\times\sum_{k=-\infty}^{\infty} \exp\left(-\frac{V\chi_t}{2}(\theta-2\pi k)^2\right)
\end{equation}
Note that the Poisson resummation (\ref{eq:Poisson}) has taken us from a sum over global topological charge
sectors, in which each term is periodic in $\theta$, to a sum over discrete $k$-vacua, where
periodicity in $\theta$ is obtained by the fact that the $k$ th term becomes the $(k+1)$ th term when $\theta\rightarrow \theta+2\pi$. 
Upon taking the logarithm to get the free energy, a single term in the sum (\ref{eq:kvacua}), with the minimum value of
$(\theta-2\pi k)^2$ dominates the logarithm in the limit $V\rightarrow\infty$. This reproduces the large N result (\ref{eq:largeN}).
We see that the purely quadratic dependence of $\varepsilon(\theta)$ which is obtained in the large $N$
approximation follows simply from the assumption of gaussian fluctuations of global topological charge.
Conversely, the deviation from quadratic $\theta$-dependence exhibited by the instanton gas model can be
related to the deviation of (\ref{eq:sum_over_instantons}) from a pure gaussian distribution. The fact that
a pure gaussian distribution in $\nu$ implies a first order transition at $\theta=\pi$ has been discussed in
\cite{Burkhalter}.

\section{Conclusions}

The results presented here demonstrate the utility of fractionally charged Wilson loops as a probe of topological structure
in $U(1)$ gauge theories. Applied to the 2D $CP^{N-1}$ models, this method clarifies the nature of the transition from
an instanton-dominated vacuum at small $N$ to a domain-wall-dominated vacuum at large $N$. Although in two dimensions both types of vacuum 
lead to confinement of fractional $U(1)$ charge, the confinement mechanisms are quite different in the two cases. This distinction
is of interest for investigating the relationship between topological charge and confinement in higher dimensional theories like $QCD$.
In two dimensions, instantons give rise to electric charge confinement by introducing a phase incoherence in the fractionally
charged loop arising from the varying number of instantons minus antiinstantons inside the loop, producing an area law falloff
of the Wilson loop. Because the instanton topological charge is locally quantized, this phase incoherence automatically disappears
for an integer charged loop. This is why the $\varepsilon(\theta)$ for the $CP^1$ model returns to zero at $\theta=2\pi$ and
exhibits no ``false vacuum'' effect. Because topological charge is not locally quantized in the large $N$, domain wall dominated 
models, the amount of charge inside the loop is not restricted to be an integer. This gives rise to phase incoherence and an 
apparently nonvanishing area law coefficient, even for integer charged loops.
As is well-known, the instanton confinement mechanism does not extend to higher dimensions. If quark
confinement in 4D QCD is to be associated with topological charge fluctuations, the instanton gas or liquid model does not
provide such a connection. On the other hand, the domain wall structure of the large $N$ $CP^{N-1}$ vacuum is
both theoretically \cite{Witten79,Luscher78} and observationally \cite{Horvath03,Ilgenfritz,Ahmad,Lian} quite analogous to
that of 4D QCD. The disordering of Wilson loop phases by layered, codimension 1 domain walls is likely to be similar in the
two cases. The study of discrete vacua and domain walls in $CP^{N-1}$ models could possibly provide insight into
quark confinement and topological phase structure in QCD. For the 2D $CP^{N-1}$ models, fractionally charged Wilson loops provide a promising
tool for studying this structure. 

This work was supported in part by the Department of Energy under grant DE-FG02-97ER41027.

\end{document}